\documentclass[conferece]{IEEEtran}
\IEEEoverridecommandlockouts
\usepackage{cite}
\usepackage{amsmath,amssymb,amsfonts}
\usepackage{algorithmic}
\usepackage{graphicx}
\usepackage{textcomp}
\usepackage{xcolor}
\usepackage{verbatim}
\usepackage{setspace}
\usepackage{makecell}
\usepackage{graphicx}
\usepackage{subfigure}
\def\BibTeX{{\rm B\kern-.05em{\sc i\kern-.025em b}\kern-.08em
    T\kern-.1667em\lower.7ex\hbox{E}\kern-.125emX}}
\usepackage[justification=centering]{caption}

\begin{document}

\title{DSRC \& C-V2X Comparison for Connected and Automated Vehicles in Different Traffic Scenarios}

\author{\IEEEauthorblockN{
Yuanzhe Jin,}
\IEEEauthorblockN{Xiangguo Liu,}
\IEEEauthorblockN{Qi Zhu}
\IEEEauthorblockA{ 
\\ECE Department, Northwestern University, Evanston, IL, USA\\
yzjin@u.northwestern.edu, xg.liu@u.northwestern.edu, qzhu@northwestern.edu}
}
\maketitle

\begin{abstract}
Researches have been devoted to making connected and automated vehicles (CAVs) faster in different traffic scenarios. By using C-V2X or DSRC communication protocol, CAVs can work more effectively. In this paper, we compare these two communication protocols on CAVs in three different traffic scenarios including ramp merging, intersection, and platoon brake. It shows there is a trade-off between communication range and interval when leveraging C-V2X or DSRC for CAVs. The result can help support further application designs for CAV autonomously choosing communication protocols in different traffic scenarios. 

\end{abstract}

\begin{IEEEkeywords}
Ramp Merging, Intersection, Connected and Automated Vehicles, Autonomous Driving, Communication Protocol, Platoon 
\end{IEEEkeywords} 
\section{Introduction}
To improve safety in Connected and Automated Vehicles(CAVs), intelligent transportation can use a new generation of communication networks and data processing capabilities to improve the overall efficiency of the existing transportation system. Among them, C-V2X and DSRC, the two communication technologies have become the first choice for applications in intelligent transportation.

Dedicated Short Range Communications(DSRC) is a wireless communication technology that uses the 5.8GHz band ISM band designed specifically for wireless communication with vehicles. As a feature of DSRC, the communication area is intentionally narrowly controlled by the directivity of the antenna and the highly accurate carrier sense. It is used in Intelligent Transport Systems (ITS), and various services are provided to drivers through communication between roadside units and on-board units\cite{6155707}. In 1999, the Federal Communications Commission (FCC) of the United States decided to allocate the 5.9GHz (5.850 - 5.925GHz) frequency band for automotive communications. The main goal is to enable public safety applications to save lives and improve traffic flow. FCC also allows the provision of private services in this field to reduce deployment costs and encourages the rapid development and adoption of DSRC technology and applications. DSRC has mature standards and good network stability. In terms of availability, DSRC has good characteristics that do not rely on network infrastructure and self-organizing networks, so the network based on the DSRC standard has strong stability and will not cause the entire system to fail due to transmission bottlenecks and single-point failures.

C-V2X is designed for low-latency direct communication and supports high-speed car scenes. The transmission mode defined by C-V2X enables a wide range of use cases. Direct C-V2X includes vehicle-to-vehicle (V2V), vehicle-to-infrastructure (V2I), etc., which can provide enhanced communication range and reliability in a dedicated spectrum. Besides, it also includes cellular networks and network communications (V2N) in the traditional mobile broadband licensed spectrum\cite{8671732}. For C-V2X, researchers apply different models to evaluate the results of semi-permanent scheduling performance communication in V2V\cite{7914699}\cite{8690990}. The C-V2X function supports both online and offline modes, and an Internet connection is no longer a necessary option. In the future, there is great potential for the development of unmanned driving.

DSRC and C-V2X, as two important communication methods on the Internet of Vehicles, have a wide range of applications. They can be used in the intelligent networked car to communicate with another vehicle explicitly. Also, the intelligent networked car can communicate with roadside units, traffic lights, road signs, and other communication methods. The purpose of vehicle-to-vehicle communication is to improve road traffic safety through functions such as lane departure warning, lane-keeping assist, and anti-collision warning.

We use Veins as the simulation software platform. Veins is an open-source framework for wireless communication simulation in a vehicle-mounted mobile environment. The underlying structure of the vehicle-mounted wireless network, such as the physical layer and the MAC layer, has been developed and perfected based on the 802.11p IEEE protocol\cite{sommer2011bidirectionally}. As a platform, Veins can simulate the performance of the vehicle in the communication process. The experiment part in III is mainly based on this for simulation.
\section{Related Work}
The contribution in this paper is mainly based on the combined research of the following aspects in ramp merging, intersection, and platoon brake.

Researchers around the world are developing various CAVs applications to solve traffic-related problems. It helps to improve the efficiency and safety of ramp merging in specific traffic scenarios, such as highways\cite{ha2020leveraging}.  In the scene of ramp merging, researchers propose a method to merge through lanes more quickly. By projecting the car from the ramp road to the straight road\cite{wang2018agentbased}. It can be noted that many of the most recent methods have adopted the control method in ramp merging. The author mentioned in the article that by projecting the cars on the ramp onto the straight. It can be approximated as a straight-line lane merging problem. This simplifies the calculation of the problem. In addition to optimization-based methods, researchers also develop other methods for ramp merging. Milanes et al. develop a logical method to allow vehicles to merge smoothly from the ramp to the mainline without changing the slope of the mainline vehicle speed to minimize the effect for crowded main line\cite{5675690}. The concept of virtual vehicle ranking by Uno et al. is used to map the virtual vehicle into the main platooning before the actual merges\cite{10.1007/978-981-13-7986-4_37}. Researchers find that driving strategies can be used to improve fuel efficiency in CAVs in different scenarios\cite{https://doi.org/10.1049/iet-its.2018.5336}, which further explains the importance of improving the efficiency of CAVs. 

Besides the research in the ramp merging, communication between the vehicles can also be helpful in other scenarios. The researchers used the Stackelberg game to model the lane-changing process of the interaction between vehicles. By making different assumptions for the two vehicles, the lanes can be switched between high and low speeds\cite{9304804}. Previous studies show that in-vehicle networks offer great hope to improve transportation safety but have strict requirements. For the intersection, with the development of autonomous driving and vehicle-to-vehicle communication technology, researchers propose a model for intelligent intersection management and developed it for analysis and simulation. It helps the design of actual vehicle network applications\cite{b6d74055e4b24cccbcb29f24f9199256}. Another research proposes a delay-tolerant protocol for intelligent intersection management, which proves the importance and effectiveness of using this framework to solve time delay in CAV network applications\cite{7946999}. In the simulation experiment, we choose this control model for intersection management.

Platoon brake is widely used in truck driving. Antonio et al. study the impact of heavy trucks operated by CAVs on highways during driving. Aiming at this specific vehicle type, it is helpful to understand the potential impact of CAVs technology on highway systems
\cite{doi:10.1061/JTEPBS.0000492}. Existing studies have pointed out that due to the limited communication range of DSRC, it cannot support the formation of a "long" platoon composed of typical trailers. To solve this problem, the researchers proposed a platoon solution based on the existing DSRC\cite{DBLP:journals/corr/abs-1910-05192}. For information exchange through wireless communication in a platoon. Researchers have studied the time-varying sampling interval, delay, and communication limitations introduced in the cooperative adaptive cruise control (CACC) system, and analyzed the stability of the system\cite{6426042}.

In \cite{9162922}, the researchers consider the vehicle network in an ideal model and show that there is a basic trade-off between these two indicators: Inter packet gap and Maximum hearing range. The Inter packet gap is defined as the time interval between sending packets and receiving packets during vehicle communication. The maximum hearing range is defined as the maximum distance that two vehicles can communicate. That is the maximum distance that a car can send a message to other vehicles and receive its feedback. In the case of high vehicle density, these two protocols need to weigh these indicators to be different. C-V2X tends to obtain more frequent communication frequencies, which means a lower inter-packet gap while DSRC tends to maintain a greater maximum hearing range. These two factors (inter-packet gap and maximum hearing range) are the key to controlling the performance of the car in the problem of different traffic scenarios like ramp merging, intersection, and platoon brake.

\section{Simulation on Traffic Scenarios}
\subsection{DSRC \& C-V2X in Ramp Merging}
We use Sumo as our simulation scenario. Sumo is a free and open-source traffic simulation suite that allows building simulations for intermodal transportation systems. Sumo includes tools that can automatically complete the core tasks for creating, executing, and evaluating traffic simulations, can be enhanced by custom models, and provide various APIs to remotely control the simulation\cite{Olaverri_Monreal_2018}. Sumo assigns appropriate routes to complete traffic demand or a group of vehicles. Its main task is to model the process of traffic participants choosing a route to their destination. Since the time to pass the edge of the road map largely depends on the number of traffic participants using this edge, route calculation is a key step to achieve large-scale traffic simulation. In Sumo, this is called user assignment or traffic assignment\cite{5625277}.

\begin{figure}[h]
\centerline{\includegraphics[width=0.5\textwidth]{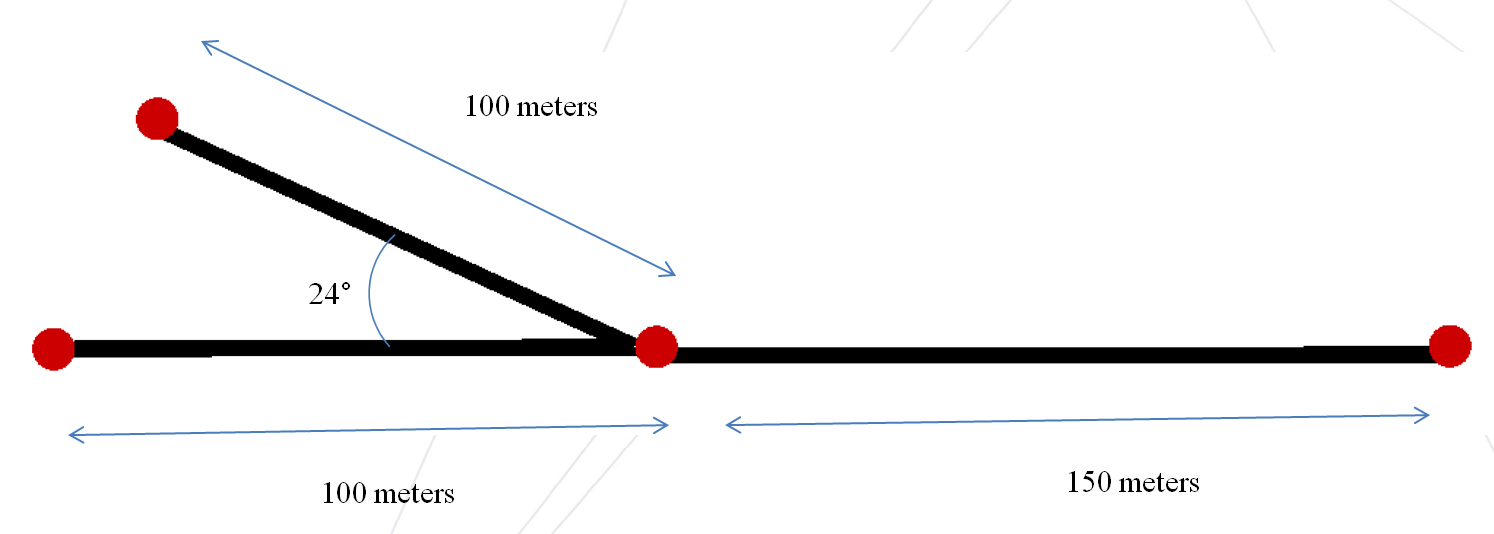}}
\caption{Ramp merging situation in the simulation experiment}
\label{ramp situation}
\end{figure}

We first build a road image in a scene where the ramps merge, as shown in Fig.\ref{ramp situation}. In this figure, the ramp merging angle at which the ramps merge is 24 degrees. The length of the road on both sides is 100m, and vehicles travel a distance of 150m after merging. We measure the time it takes for vehicles to pass through the same road section under different communication protocols to compare the impact of vehicle passing time caused by different communication protocols under different traffic densities.

Sumo can be regarded as a pure microscopic traffic simulation. Each vehicle is given an identifier, time of departure, and route of the vehicle in the road network. A macro traffic simulator treats the entire traffic flow as a unit. Sumo can also define departure and arrival attributes such as lane, speed, or location. Each vehicle is assigned a type, which describes the physical characteristics of the vehicle and the variables of the motion model\cite{SUMO2018}.

The simulation is discrete in time and continuous in space, and internally describes the location of each vehicle, that is, the lane and the distance from the starting point. When the vehicle is moving, the following model is used to calculate the vehicle speed. Sumo's Car-Driver Model can show the main characteristics of the traffic. At each time step, the speed of the vehicle is adapted to the speed of the preceding vehicle, avoiding collisions in subsequent steps. This speed is called Safe Velocity, and the formula for it is in Equation \ref{safety speed}.

\begin{equation}
v_{s a f e}(\mathrm{t})=v_{l}(t)+\frac{g(t)-v_{l}(t) \tau}{\frac{\bar{v}}{b(\bar{v})}+\tau}
\label{safety speed}
\end{equation}

In Equation \ref{safety speed}, ${v_{l}(t)}$ is the speed of the preceding vehicle, g(t) is the distance between the preceding vehicle, $\tau$ is the driver’s reaction time, and b(v) is the deceleration function.

The desired speed (${v_{des}(t)}$) of the vehicle takes the minimum of the following three: the safety velocity calculated above, the vehicle speed plus the maximum acceleration, and the maximum possible speed. Therefore, its expected speed is shown in Equation \ref{desire speed}.
\begin{equation}
v_{d e s}(t)=\min \left[v_{s a f e}(t), v(t)+a, v_{\max }\right]
\label{desire speed}
\end{equation}

The desired speed ensures that vehicles should not collide with each other in the simulated environment. This constraint ensures that no matter what traffic density conditions, the process of vehicle driving is safe.

This speed is consistent with the time-varying speed of the ramp merging proposed in the paper\cite{wang2018agentbased}. Under these guaranteed conditions, we can change the maximum hearing range and the communication frequency of the vehicle in the Veins simulation environment. Then we can calculate the time for the same vehicle to merge into the straight through the same road under different conditions and compare the result.

The angle between the straight road and the merging ramp affects the communication distance between vehicles. In the previous experiment, the ramp road shown in Fig.\ref{ramp situation} that the experiment used has an angle of 24 degrees, which means the maximum distance between the straight line and the ramp is less than the minimum distance that C-V2X can communicate. C-V2X can communicate without any effect.

However, if the angle of the ramp reaches or exceeds a certain degree (in some extreme situations like intersections), we can not ignore the influence of the distance between the vehicles on the communication of the vehicles. Among the two communication methods mentioned in this paper, DSRC maintains the same communication distance when the vehicle density exceeds 250 vehicles. Under the condition that the number of vehicles is greater than 250, the influence of the angle between the ramps on the communication of the C-V2X method can not be ignorant. So we design the experiment to compare the efficiency of C-V2X in the different ramp merging angles. We test it in the ramp merging angles of 24, 48, and 72 degrees.

\subsection{DSRC and C-V2X in Intersection}
\begin{figure}[h]
\centerline{\includegraphics[width=0.45\textwidth]{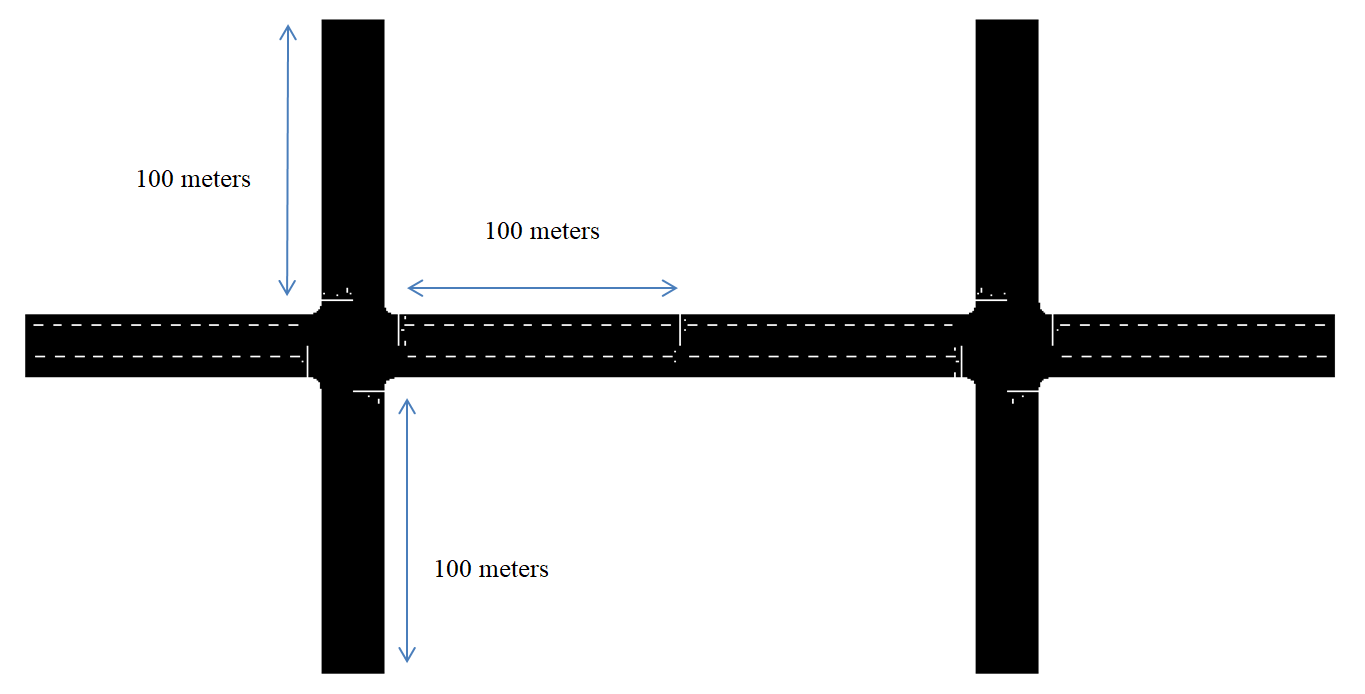}}
\caption{Intersection situation in the simulation experiment}
\label{intersection situation}
\end{figure}

Researchers have proposed a delay-tolerant protocol for intelligent junction management of communication delays and developed a model for analyzing the performance of the protocol
\cite{b6d74055e4b24cccbcb29f24f9199256}. With the help of related models developed by it, we design two same connected intersections with each direction in 100 meters, which is shown in Fig.\ref{intersection situation}. The right intersection is the same as the left one. And all the sides of the intersection are set to be 100 meters to simplify time measurement.

The intersection control method is used from an intelligent intersection management model\cite{b6d74055e4b24cccbcb29f24f9199256} which can control CAVs pass through the intersection without deadlock. The model can set the communication distance as the maximum hearing range for the communication protocol and it also has the parameter as the communication interval corresponding to the inter-packet gap. The CAVs can be generated randomly from each direction from the intersection to pass through it. And we can test the efficiency of DSRC and C-V2X by measuring the total time for all the generated CAVs passing through the intersection.

\subsection{DSRC \& C-V2X in Platoon Brake}
Platoon Brake is a typical situation in CAV cruises. The vehicles need to stop due to some emergencies or wait for a red light at the intersection. At this time, the head vehicle brakes, and the subsequent vehicles stop with the head. In the end, the entire platoon can be parked while maintaining the same inter-vehicle distance(IVD). Plexe, as an open-source extension of Veins, provides researchers with a simulation environment that can run experiments in real-world scenarios\cite{7013309}. In the platoon brake part of the experiment, we use Plexe to simulate the performance of DSRC and C-V2X. The platoon brake situation is simulated in a 10 km straight road with different amounts of vehicles in the platoon. The vehicle model is simplified as a 20-meter rectangle to simulate a truck. And the inter-vehicle distance is set to be 10 meters at every simulation.

In the experiment, we use some definitions to measure the performance of the communication protocol in platoon brake. The Brake Distance is used as the distance for the whole platoon to stop. Platoon Speed is defined as the cruising speed for the platoon. The main measurement for the platoon brake is to use Brake Time(B-Time). B-Time is the time for the whole platoon to stop from the head vehicle to brake until the last vehicle stop. And we also add another measurement as Min Inter-vehicle Distance(MIVD), which is defined as the minimum distance between the vehicles in the platoon when the platoon stops. 
\section{Result \& Conclusion}
\subsection{DSRC \& C-V2X in Ramp Merging}
After the simulation in Veins, the results are shown in Table \ref{comparison}. We take road time as the standard for comparing the difference between the two communication protocols. Road time is defined as the time of the same CAV that begins from its appearance in the road and ends at their merging into the straight line. Through Table \ref{comparison}, it can be found that when the traffic density is not large, the road time of the two communication protocols has little difference. With the increase in traffic density, the C-V2X has a better performance than DSRC.
\begin{table}[h]
\centering
\begin{tabular}{c|c|c|c|c}
\hline
Protocol & V-Density & MHR(km) & IPG(ms) & Times(s) \\ \hline
C-V2X      & 250          & 0.200                    & 100                   & 12.14          \\ \hline
C-V2X      & 750          & 0.066                    & 100                   & 13.03          \\ \hline
C-V2X      & 1500         & 0.033                    & 100                   & 24.63          \\ \hline
C-V2X      & 2000         & 0.025                    & 100                   & 38.11          \\ \hline
DSRC       & 250          & 0.250                    & 125                   & 12.03          \\ \hline
DSRC       & 750          & 0.250                    & 375                   & 13.15          \\ \hline
DSRC       & 1500         & 0.250                    & 750                   & 26.58          \\ \hline
DSRC       & 2000         & 0.250                    & 1000                  & 40.51          \\ \hline
SPE*       & 250          & 0.300                    & 100                   & 12.01          \\ \hline
SPE*       & 250          & 0.010                    & 1000                  & 13.97          \\ \hline
SPE*       & 2000         & 0.010                    & 1000                  & 42.17          \\ \hline
\end{tabular}
\caption{A comparison between C-V2X and DSRC in ramp merging.V-Density is short for Vehicle Density; MHR is short for Maximum Hearing Range and IPG is short for Inter-packet Gap.}
\label{comparison}
\end{table}

Except for the influence of the communication protocol itself, under the same communication protocol, the time that CAV travels on the same road is related to the inter-packet gap and the maximum hearing range. The result in Table ~\ref{comparison} also confirms the theoretical results\cite{9162922}. The time for a car to pass through the ramp merging situation increases as the inter-packet gap increases. And the time for a car to pass through the road increases as the maximum hearing range decreases. It is clear to understand because as the vehicle density increase, CAVs take more time to merge and the communication between CAVs takes more time. We also have some experiments showing some extreme situations. When the inter-packet gap and the maximum hearing rage are not a trade-off. In Table~\ref{comparison}, the SPE* is a fictitious communication protocol, which has a shorter maximum hearing range and a longer inter-packet gap. And it is neither as good as DSRC nor as C-V2X. 

\begin{figure}[]
\centerline{\includegraphics[width=0.35\textwidth]{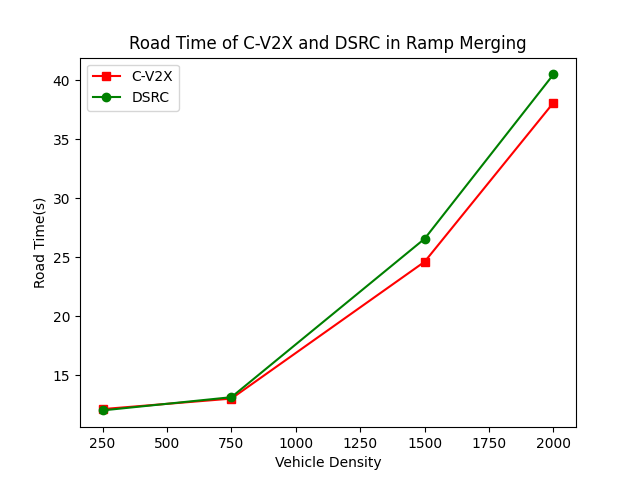}}
\caption{The road time consumption between C-V2X and DSRC in the ramp merging}
\label{road time}
\end{figure}
In Fig.\ref{road time}, it shows the relation between road time and vehicle density from Table~\ref{comparison}. For different traffic conditions, if we put these results in a line chart like Fig.~\ref{road time}, it can be found that C-V2X performs better than DSRC in high vehicle density and both perform similarly in low car density.

For different ramp merging angles, we only make the experiment on C-V2X because DSRC keeps the maximum hearing range constant under certain vehicle density conditions. Considering when the vehicle density is small, the communication simulation of different communication protocols under the condition of 24° included angle has been completed. Therefore, this experiment is simplified to the effect of the angle on-ramp merging using the C-V2X protocol. As the inter-packet gap of vehicle communication is not considered, this experiment is also suitable for DSRC when the traffic density is small when the hearing range is the only variant.

\begin{figure}[h]
    \centering
    \subfigure[]{
        \includegraphics[width=0.22\textwidth]{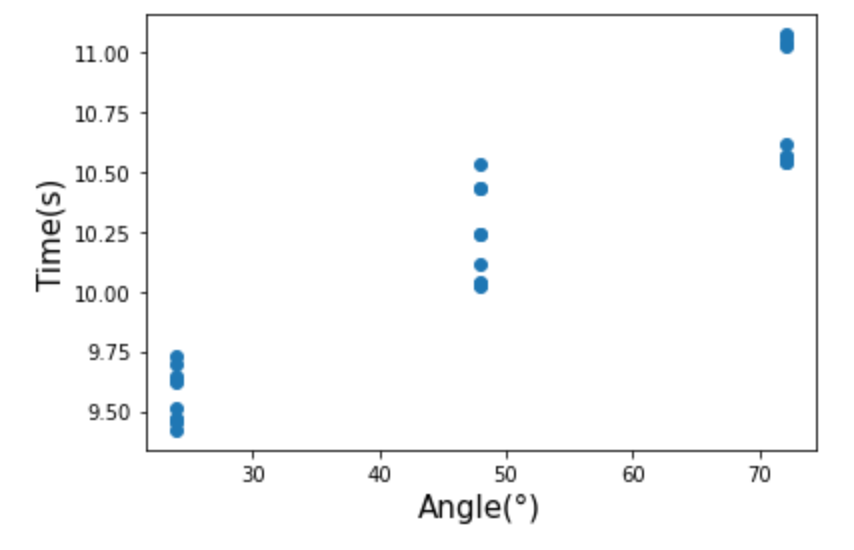}}
    \centering
    \subfigure[]{
        \includegraphics[width=0.22\textwidth]{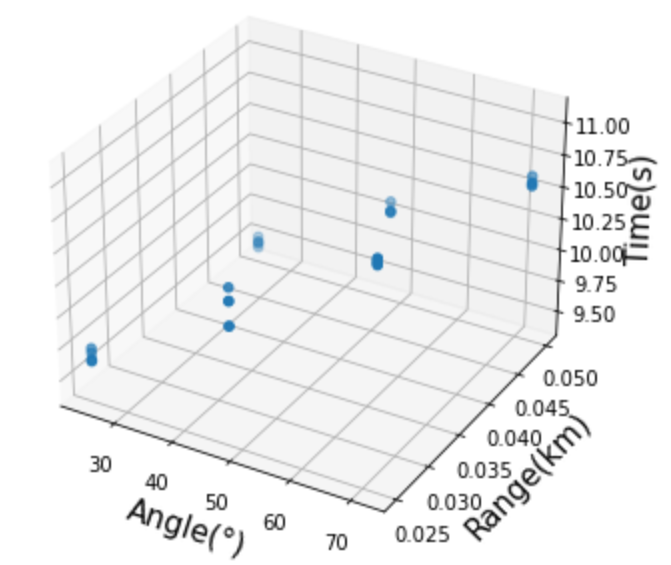}}
    \caption{(a) The relation between road time, merging angle and hearing range. (b) The relation between the merging time and the merging angle}
\label{A-R-T}
\end{figure}

In the experiment, we also set road time (Time) or merging time as a standard to measure the efficiency of the communication protocol for CAVs in ramp merging situations, which is similar to the definition in the same ramp merging angle situation. It measures the time a vehicle appears from the ramp start point and merges into the straight road. The result between the ramp merging angle and road time is shown in Fig.\ref{A-R-T}(a). In Fig.\ref{A-R-T}(a), the relation between road time, ramp merging angle shows that as the angle becomes larger, the time for a vehicle merging into the road becomes longer.
The angle affects the road time indirectly by changing the hearing range of a vehicle. If we put road time, ramp merging angle, and hearing range into the same figure, it shows that the road time has a positive relation with ramp merging angle and hearing range. In Fig.\ref{A-R-T}(b), the road time increases as the ramp merging angle and hearing range increases. Through the relationship between ramp merging angle and hearing range, it shows there is a positive correlation between road time and ramp merging angle.

\subsection{DSRC \& C-V2X in the Intersection}
\begin{table}[h]
\centering
\begin{tabular}{c|c|c|c}
\hline
Protocol & V-Density & MHR(km) & Time(s) \\ \hline
C-V2X    & 700             & 0.07    & 68.35   \\ \hline
C-V2X    & 550             & 0.09    & 65.01   \\ \hline
C-V2X    & 400             & 0.13    & 56.76   \\ \hline
C-V2X    & 250             & 0.20    & 38.54  \\ \hline
C-V2X    & 100             & 0.50    & 30.60   \\ \hline
DSRC     & 700             & 0.25    & 59.77   \\ \hline
DSRC     & 550             & 0.25    & 56.46   \\ \hline
DSRC     & 400             & 0.25    & 52.55   \\ \hline
DSRC     & 250             & 0.25    & 38.32   \\ \hline
DSRC     & 100             & 0.50    & 30.60   \\ \hline
\end{tabular}
\caption{C-V2X and DSRC in Intersection}
\label{intersection result table}
\end{table}

Table \ref{intersection result table} shows the performance of C-V2X and DSRC in different vehicle density situations. Compared to the same vehicle density, C-V2X performs better than DSRC. And the table also shows that as the maximum hearing range increases or the inter-packet gap decreases, the time for CAVs to pass the intersection becomes shorter.

\begin{figure}[h]
\centerline{\includegraphics[width=0.35\textwidth]{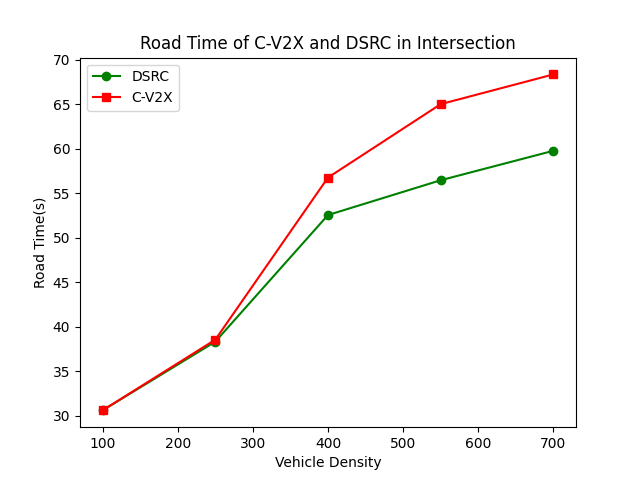}}
\caption{A comparison between C-V2X and DSRC performances in Intersection}
\label{intersection result figure}
\end{figure}

We can draw the Fig.\ref{intersection result figure} from the Table \ref{intersection result table}, which shows DSRC and C-V2X has an approximate result in low vehicle density but DSRC performs better in the high vehicle density when the number of vehicles exceeds 250 cars per hour. It meets the previous study that when the density of the vehicle is higher than 250, the difference between C-V2X and DSRC begins.DSRC makes the vehicle pass the intersection in less time than C-V2X.

\subsection{DSRC \& C-V2X in Platoon Brake}
\begin{table}[h]
\centering
\begin{tabular}{c|c|c|c}
\hline
Protocol & V-Density & Brake Time(s) & MIVD (m) \\ \hline
C-V2X    & 250       & 3.43          & 8.96     \\ \hline
C-V2X    & 500       & 3.51          & 8.93     \\ \hline
C-V2X    & 1000      & 3.74          & 8.86     \\ \hline
C-V2X    & 1250      & 3.82          & 8.81     \\ \hline
C-V2X    & 1500      & 4.13          & 8.74     \\ \hline
DSRC     & 250       & 3.68          & 8.89     \\ \hline
DSRC     & 500       & 3.95          & 8.13     \\ \hline
DSRC     & 1000      & 4.52          & 6.61     \\ \hline
DSRC     & 1250      & 4.72          & 5.84     \\ \hline
DSRC     & 1500      & 4.81          & 5.09     \\ \hline
\end{tabular}
\caption{-V2X and DSRC in Platoon Brake}
\label{platoon brake table}
\end{table}

In the experiment, the platoon needs to maintain the same inter-vehicle distance. Every time the platoon brakes, the lead car is the first one to brake. So the brake distance remains the same if we set the platoon at a certain speed and inter-vehicle distance. Brake time and MIVD are used in the conclusion to show the performance between C-V2X and DSRC in the platoon brake situation.

\begin{figure}[h]
    \centering
    \subfigure[]{
        \includegraphics[width=0.22\textwidth]{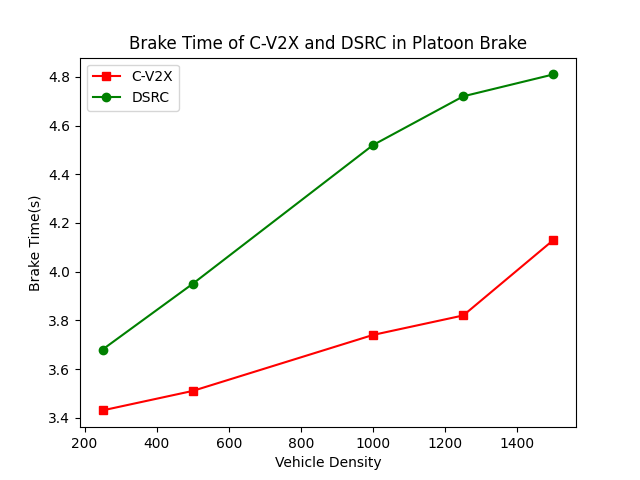}}
    \centering
    \subfigure[]{
        \includegraphics[width=0.22\textwidth]{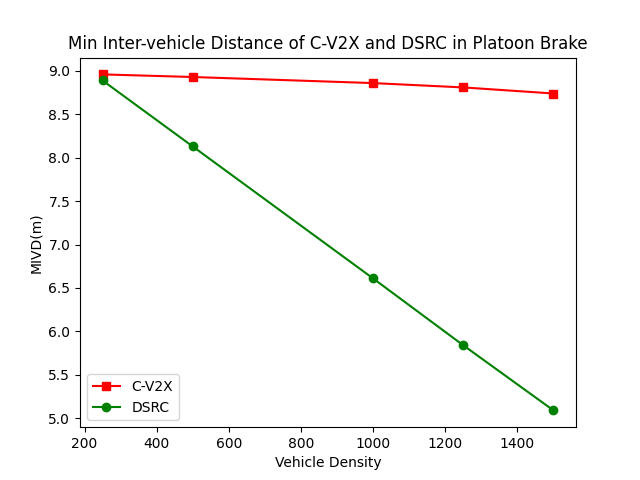}}
    \caption{(a) Brake time between C-V2X and DSRC. (b) MIVD between C-V2X and DSRC}
\label{platoon brake}
\end{figure}

For brake time, it can be observed in Table \ref{platoon brake table} that C-V2X performance better than DSRC. It has a slight difference in low vehicle density but becomes large when the vehicle density increases. In Fig.\ref{platoon brake}(a), it shows that the platoon needs more time to brake using DSRC compared to C-V2X. For MIVD, it can be observed in the experiment that when the IVD is set as a constant, the MIVD becomes shorter as the inter-packet gap becomes longer during braking. In other words, a short MIVD makes the platoon unsafe. From Table \ref{platoon brake table} and Fig.\ref{platoon brake}(b), the vehicle density becomes longer, MIVD becomes shorter. And C-V2X has little changes in MIVD as vehicle density changes. But for DSRC, the MIVD drops as the vehicle density increases. 

\section{Conclusion}
For each traffic situation, we can get a possible explanation for the performance between DSRC and C-V2X.

For the same ramp road (that is, the impact of ramp merging angle on the communication between CAVs is not considered), the possible explanation for these results is that in low car density, with low inter-packet gap and large maximum hearing range, C-V2X and DSRC perform almost the same in the ramp road. When the vehicle density is low, the vehicle has a suitable time to merge, and the high communication frequency wastes a certain amount of time. The trade-off makes the result approximately the same.

In high car density, the possible explanation is that although the C-V2X does not communicate with each other when the vehicles entered the road with a long distance, the frequent communication between the vehicles when approaching the merging point ensures that the vehicles can be merged into the platoon at an appropriate time. For DSRC, though vehicles can start communicating with each other when vehicles just enter the road, due to the long inter-packet gap of communication between vehicles, the vehicle may miss the appropriate merging time while waiting for communication. So it needs to wait for the next appropriate time to merge. If we take the ramp merging angle into consideration. Generally, when there are no obstacles that hinder signal propagation on the road, the ramp merging angle only affects the communication distance of CAVs. The ramp merging angle indirectly affects the distance of communication transmission and affects the traffic efficiency of C-V2X. As the ramp merging angle becomes larger, the distance between two CAVs becomes longer, and it makes the inter-packet gap longer. It finally affects communication and makes the CAVs less efficient.

In the intersection situation, we discover that the result between DSRC and C-V2X in the low vehicle density is similar. While in the high vehicle density, DSRC performs a bit better than C-V2X due to its long communication range between the vehicles. When the vehicle arrives at the intersection in a long-distance, the infrastructure can ensure the vehicles can pass it at an appropriate time. In an ideal situation, the intersection control method can allocate the proper time for the vehicle for only one-time communication. It makes DSRC performances better than C-V2X because it is no need to often communicate with the infrastructure.

For the platoon brake situation, a short MIVD means the two vehicles are very close to each other, which is not considered to be safe. In an ideal situation, MIVD can be equal to inter-vehicle distance. Due to the communication interval, the following vehicles cannot stop as soon as the leader vehicle stops, which makes MIVD less than the inter-vehicle distance. In both brake time and MIVD, C-V2X performs better than DSRC. The possible explanation is that vehicles in the platoon send the message one after the other by their position. It affects DSRC more than C-V2X. And C-V2X performance better than DSRC in platoon brake.

According to the result from the experiment, it can help researchers design some applications using machine learning methods for CAVs which can make CAVs autonomously choose their communication protocol in different traffic scenarios. The application can enable CAVs to choose their proper communication protocols by detecting the surrounding of them. The future improvements of the work also include testing more different scenarios between the two communication methods in more cases. It has a great significance to improve the efficiency of CAVs. Not only can it save ordinary people's usual commute time, but it also has a great impact on reducing carbon emissions. Other CAV situations such as highways, roundabouts, etc. would be helpful to test the relationship on how the maximum hearing range and the inter-packet gap can affect CAVs in practical use.

\bibliographystyle{IEEEtran}
\bibliography{ref}

\end{document}